\documentclass[floatfix,reprint,twocolumn,notitlepage,prb,superscriptaddress,showpacs]{revtex4-2}
\usepackage{hyperref}
\usepackage[usenames,dvipsnames]{color}
\usepackage{amsmath}
\usepackage[english]{babel}
\usepackage{amssymb}
\usepackage{graphicx}
\usepackage{epstopdf}
\usepackage{comment}
\usepackage{soul}
\usepackage{marvosym}
\usepackage{mathrsfs}
\usepackage{amsfonts}
\usepackage{tabularx}
\usepackage{multirow}

\let\oldbibitem\bibitem
\renewcommand{\bibitem}{%
  \renewcommand{\doi}[1]{doi: ##1}
  \let\bibitem\oldbibitem
  \oldbibitem
}

\begin{document}

\title{Contact Angle Studies on Porous Silicon: Evidence for Heterogeneous Wetting and Implications of
Oxidation}
\author{S. J. Spencer}
\author{C. G. Deacon}
\author{G. T. Andrews}
\email{tandrews@mun.ca}
\affiliation{Department of Physics and Physical Oceanography, Memorial University of Newfoundland and Labrador, St. John's, Newfoundland \& Labrador, Canada A1B 3X7}

\date{\today}
\begin{abstract}
A study of wetting was carried out on porous silicon films with pore diameters spanning three orders of magnitude. Water contact angle measurements on adjoining porous and nonporous regions yielded Wenzel roughness ratios that were either unphysical (less than unity) or unrealistically low when compared to those expected from specific surface area considerations. Moreover, results obtained from a sample consisting of a microporous film on a macroporous layer gave contact angles that were very similar to those found on films consisting of only a single microporous layer, contrary to what would be expected for complete filling of the pores by liquid. Values for wetted surface fractions calculated from the Cassie-Baxter model are unreasonably high for micro- , meso- and oxidized macroporous films considering their porosities, while relatively oxide-free macroporous films give wetted surface fractions in accord with those expected based on film porosity. Collectively, these results show that the predominant mode of wetting on these films is heterogeneous. 
\end{abstract}

\pacs{68.08.-p, 68.08.Bc, 68.03.Cd, 61.66.Bi, 72.80.Cw, 81.20.-n}

\maketitle

\section{Introduction}
Wetting in porous media is a subject of current research interest. From a fundamental perspective, the motivation for studying this phenomenon lies in revealing the mechanisms responsible for the mode of wetting - homogeneous or heterogeneous. This knowledge is important because it permits control of the wetting state for applications in areas such as biofouling \cite{marm2006} and sensing \cite{kova2009}, via alteration of the physical and/or chemical properties of the pore network and associated surface. Porous silicon is an especially useful platform in this regard because of the flexibility it provides in terms of pore geometry and morphology, and the ease with which the surface may be altered by chemical treatment.  

The two most familiar models of wetting that are applied to rough and/or porous media are those of Wenzel \cite{wenz1936} and Cassie-Baxter \cite{cass1944}.  In the Wenzel model, the contact angle on a rough surface, $\theta_W$, is related to that on a planar surface of the same material, $\theta_Y$ (so-called Young's angle), via 
\begin{equation}
\cos \theta_W = r_W \cos \theta_Y,
\end{equation} 
where $r_W$ is the areal ratio of structured surface to planar surface \cite{naga2009}. In the Wenzel model, the liquid wets the entirety of the textured surface under the drop.  This is referred to as homogeneous wetting. 

The Cassie-Baxter model applies to the specific case of heterogeneous wetting in which the liquid does not enter the pores. It relates the fraction of solid surface area wet by the liquid ($f$) to the contact angle, $\theta_{CB}$, according to \cite{cass1944,marm2003,borm2010,shim2014}
\begin{equation}
\cos \theta_{CB} = f\cos \theta_Y + f - 1.
\end{equation}
Models have also been developed to describe heterogeneous wetting in which the liquid partially infiltrates the pores \cite{naga2009,borm2010}.

Most studies of wetting on porous silicon have focused on adjustment of the hydrophobicity by functionalization \cite{naga2009,dest2008,liu2009,coff2013,wojt2002,jarv2012,vyhm2013,wu2012,datt2006,datt2007,khun2007,ress2008,bjor2006} or modification of the pore geometry \cite{naga2009,ress2008,cao2008,nova2011}, while only a few investigate the applicability of wetting models \cite{naga2009,nova2011,spen2013,nenz2011}. Two studies have applied the Wenzel model to mesoporous silicon \cite{nova2011,spen2013}, while two other studies on macroporous films suggests that neither the Wenzel model nor the Cassie-Baxter model apply \cite{naga2009,nenz2011}. In this paper, a study of wetting on micro-, meso- and macroporous silicon films is reported. The results show that heterogeneous wetting is the predominant mode of wetting on porous silicon regardless of pore diameter.

\section{Experimental Details} \label{sec:exp_details}
\begin{table*}
\caption{Substrate resistivity, preparation conditions and resulting porosity and estimated thickness for porous silicon films.}
\begin{ruledtabular}
\begin{tabular}{ccccccc}
Type &Resistivity &Current Density  & Etch Time &Electrolyte & Porosity & Thickness\\
 &($\Omega$$\cdot$cm) &(mA$\cdot$cm$^{-2}$)      & (s) & & (\%) & ($\mu$m) \\ \hline
Microporous   &   2.5 - 4.0             &  10.7     &  900     &9 (48$\%$) HF:1 C$_{2}$H$_{5}$OH & 60 & 8 \\
Mesoporous   &   0.005 - 0.02       &  164.3   &  33.8   &1 (48$\%$) HF:1 C$_{2}$H$_{5}$OH & 60 & 5 \\
Macroporous  &  9.0 - 13.0           &  10.0      &  600    &1 (48$\%$) HF: 9 ACN & $\sim 40$ & 1.5 \\ 
\end{tabular}
\end{ruledtabular}
\label{tab:samp_prep}
\end{table*}

\subsection{Sample Preparation}
Micro-, meso-, and macroporous silicon samples were formed via electrochemical etching of (100)-oriented crystalline silicon in a PTFE cell at constant current under the conditions shown in Table \ref{tab:samp_prep}. All samples were given a 60 s HF bath prior to film formation to remove native oxide present on the surface. Micro- and mesoporous samples were prepared using recipes that have been used in previous work which yield a porosity of 60\% for both \cite{pars2012,polo2012}.
The microporous film thickness was estimated by substituting the etch time and current density into the equation $dh/dt = 1.05J^{0.89}$, where $dh/dt$ is the etch rate and $J$ is current density \cite{lehm2002}. The thickness of the mesoporous films was estimated from the etch time and experimentally-determined etch rates \cite{polo2012}. Macroporous silicon films were formed using a process similar to that described in Ref. \cite{peck2013}. A microporous layer is formed on top of the macroporous
film and this layer was removed on all but one sample by etching in a 1\% KOH solution (see Fig. 1).  The macroporous samples were then given another 60 s HF bath to remove the surface oxide created by the KOH etch \cite{alv2008}. The macroporous film thickness was measured from cross-sectional SEM images. In addition to the porous region, it is important to note that each sample has an adjoining area that was not exposed to electrolyte and hence is nonporous. Aside from the electrochemical etching itself, these areas were subjected to the same processing and environmental conditions as the corresponding porous regions.  After fabrication, all samples were stored in vacuum for 24 hours prior to contact angle measurement.

\begin{figure}
\includegraphics[scale=0.2]{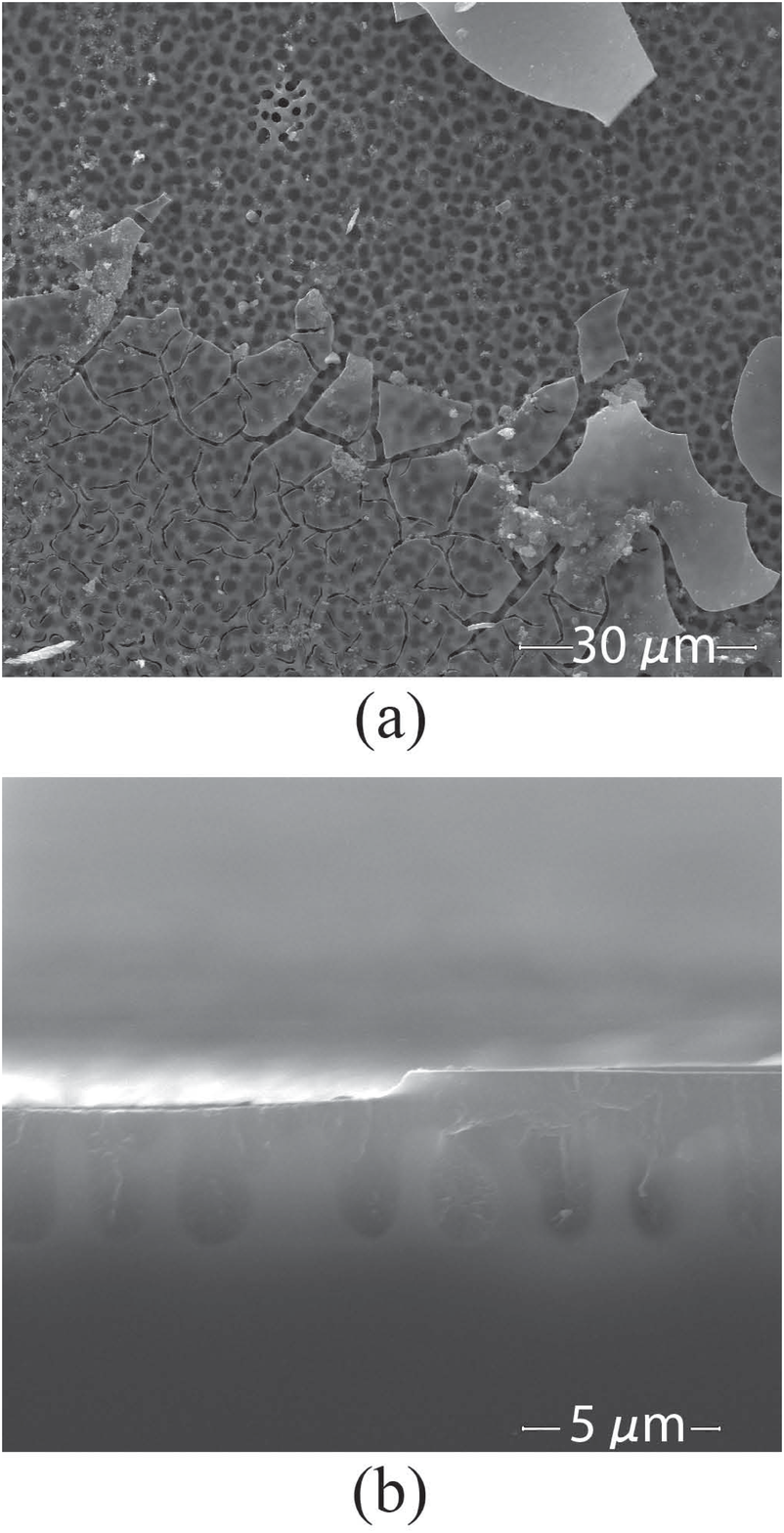}
\caption{Scanning electron microscopy images of microporous layer on macroporous silicon film: (a) plan view of macroporous silicon film with micorporous layer present on the lower left and removed via etching in 1\% KOH solution on the upper right, (b) cross-sectional view with microporous layer present on right and removed via etching in 1\% KOH solution on the left. }
\label{fig:micro_on_macro}
\end{figure}

\begin{figure}
\begin{center}
\includegraphics[scale=0.184]{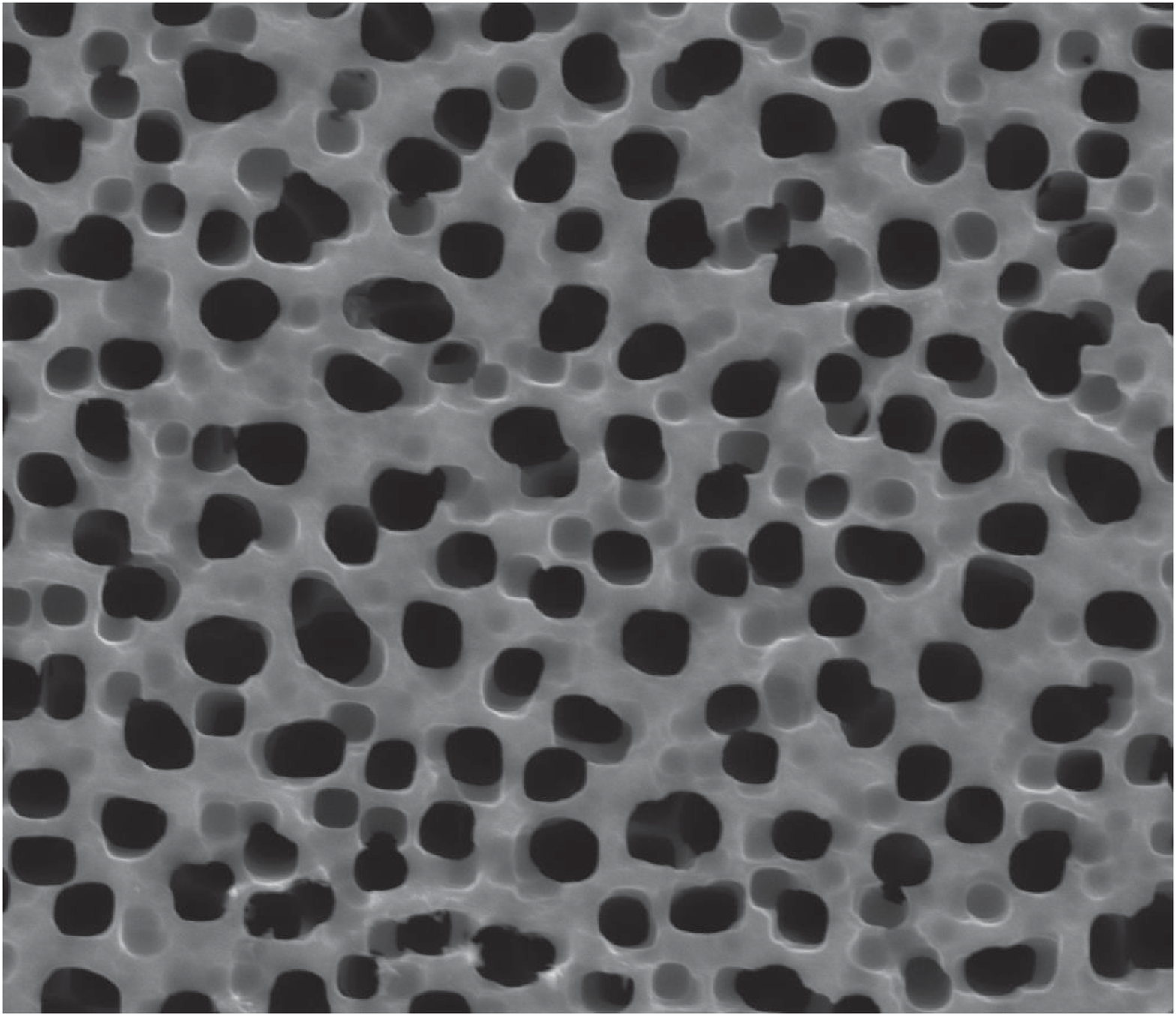}
\includegraphics[scale=0.42]{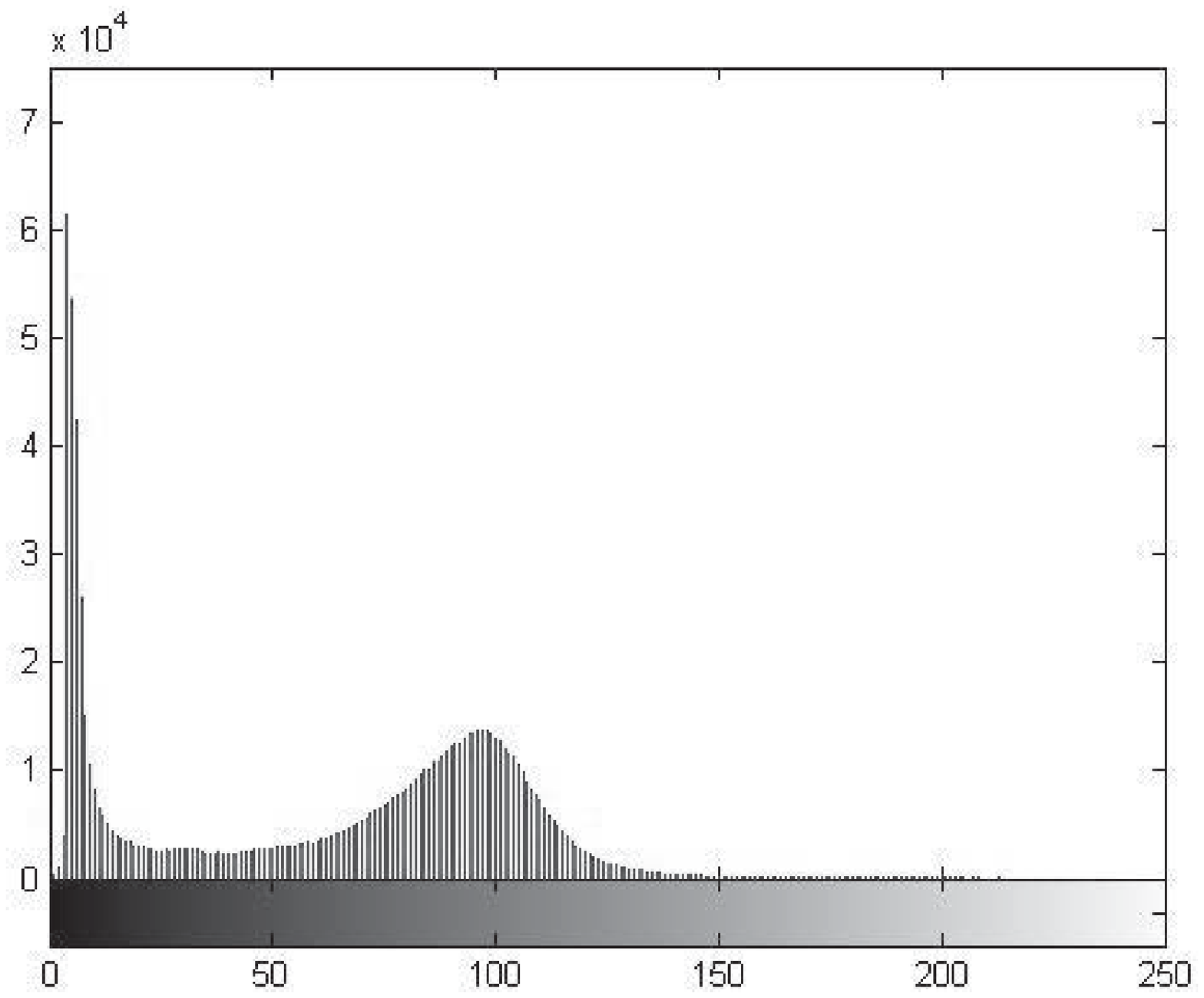}
\caption{(a) Scanning electron micrograph of a macroporous silicon film saved as a greyscale image. (b) Histogram showing greyscale distribution obtained from the image shown in (a). The bins  corresponding  to the pores lie to the left of channel 50.  The porosity was determined to be $\sim 40\%$.}
\label{fig:hist}
\end{center}
\end{figure}

\subsection{Determination of Macroporous Film Porosity} \label{det_macro_film_porosity}
The porosity of the macroporous silicon films was estimated from SEM images using a Matlab algorithm. SEM micrographs of $\sim 10$ samples prepared in the same manner as those used for contact angle measurements were saved as greyscale images with each pixel in an image assigned an integer grey value between 0 (black) and 255 (white). A histogram showing the greyscale distribution was produced for each image. Typical results are shown in Figure 2.  Each histogram showed two distinct peaks: one near the lowest numbered bins corresponding to the pores (which appear black in the micrographs), and the second near channel 100 corresponding to the light greys of the film surface. Channel 50 was assigned as the cut-off point between these two regions. The areal porosity was determined by dividing the total number of pixels below the cut-off channel by
the total number of pixels in the image. Due to the columnar nature of the pores and the planarity of the interface between the porous film and the nonporous substrate (see Fig. 1(b)), the volumetric porosity is approximately equal to the areal porosity. The average porosity for the macroporous films was found
in this way to be $\sim 40$\%.

\subsection{Contact Angle Measurements}
Water contact angles for several drops were measured on both the porous and nonporous (crystalline) regions of each sample using the procedure detailed in Ref. \cite{spen2013}.

\begin{table*}
\caption{Average contact angles measured on micro-, meso-, and macroporous silicon films and adjoining nonporous regions. $f$ corresponds to the wetted surface fraction obtained from the  Cassie-Baxter equation. The numerical quantity in parentheses in the `Type of Film' column is the approximate pore diameter and that in the `Contact Angle' columns is the standard deviation.  Entries marked with an asterisk $(*)$ indicate films for which contact angle measurements yield physical values for $r_W$. } \label{tab:ave_contact_angle}
\begin{ruledtabular}
\begin{tabular}{cccccc}
Type of Film                                                        &   \# Drops        &   Contact Angle   &  \#  Drops  & Contact Angle    &  $f$   \\ 
                                                                           &    (Porous)  & (Porous)           &  (Nonporous)   &  (Nonporous)    &         \\ \hline
\multirow{7}[4]{3 cm}{Microporous \\ ($\sim 1$ nm)  }    &      3     &   81 (2)          &  3 &      77 (3)              & 0.94   \\
                                                               &     2     &     84 (8)       & 3  &         89 (1)              & $\ast$    \\
                                                                          &     3      &      87 (2)         & 3 &        75 (5)              & 0.84    \\
                                                                         &      3       &      $\;\,$86 (11)     & 3  &         72 (4)              & 0.83   \\ 
                                                                         &      3       &      82 (3)        &  3 &       71 (2)             & 0.86  \\
                                                                         &     3       &      77 (5)         & 3 &         70 (5)             & 0.91  \\
                                                                         &     3       &      76 (7)         & 3 &         70 (3)             & 0.93   \\ \hline
\multirow{6}[4]{3 cm}{Mesoporous \\ ($\sim 10$ nm) } &   3        &      71 (1)         &  3 &        79 (5)             & $\ast$       \\
                                                                        &    2       &      $\;\,$79 (18)      & 3   &          79 (9)            & 1.00        \\
                                                                        &    3        &      77 (4)       & 2  &          71 (1)            & 0.92   \\
                                                                         &   3        &      77 (2)        & 3 &          73 (4)            & 0.95   \\
                                                                         &  3         &      73 (8)         & 3 &          62 (1)            & 0.88  \\
                                                                        &  3          &      $\;\,$69 (10)    & 3   &          59 (5)            & 0.90  \\  \hline       
\multirow{5}[4]{3 cm}{Macroporous (As-Prepared) ($\sim 1\ \mu$m)} &      3     &       49 (5)        &  3 &        44 (7)          & 0.97   \\
                                                                        &  2         &       $\;\,$53 (10)    &  3  &          $\;\,$51 (16)         & 0.98  \\
                                                                         &  3          &       75 (7)        &  3 &        $\;\,$42 (15)         & 0.73  \\
                                                                         &  2          &       67 (4)        &  3 &       $\;\,$46 (15)         & 0.82  \\
                                                                         &  3          &       $\;\,$70 (15)        &  3 &        36 (3)           & 0.74   \\ \hline
\multirow{4}[4]{3 cm}{Macroporous (HF-Dipped) \\ ($\sim 1\ \mu$m)}   &      3      &       $\!\!\!$101 (6)      & 2 &         66 (7)           & 0.57    \\
                                                                        &   3       &        99 (3)        &  3 &        64 (5)          & 0.58   \\
                                                                         &  3        &       $\!\!\!$111 (4)       & 3 &          69 (2)          & 0.47   \\
                                                                         &  3        &       $\!\!\!$100 (7)       & 3 &         66 (4)          & 0.59   \\  \hline
Micro on Macro (As-Prepared)  &       3     &        83 (3)        &           3                   &  55 (3) &     0.71        \\ \hline
Micro on Macro (HF-Dipped)     &       1     &        99 (3)         &            1                   & 67 (3) &   0.61              \\
\end{tabular}
\end{ruledtabular}
\label{tab:ave_cont_ang}
\end{table*}

\section{Results and Discussion}
\subsection{Contact Angle: Nonporous Regions} \label{sec:cont_ang_nonporous}
Table \ref{tab:ave_cont_ang} shows contact angle data for all films studied. Contact angles on the nonporous regions of the microporous and mesoporous samples are, in general, slightly less than that for H-terminated crystalline silicon ($\sim 80^\circ$) \cite{naga2009,coff2013,nova2011}. This is not surprising because the last treatment that this part of the samples received is a pre-etch HF dip which leaves the surface H-terminated \cite{laue1993}, followed by a water rinse. The reduction in angle relative to H-terminated silicon is likely due to some degree of surface oxidation as a result of ambient air exposure.  In contrast, the nonporous region of the as-prepared macroporous sample has a contact angle comparable to that for water on fully oxidized crystalline silicon ($\sim 44^\circ$) \cite{naga2009,spen2013}. Again, this is expected because immersion in KOH (to remove the microporous layer that lies on top of the macroporous film) results in oxidation \cite{alv2008} which, in turn, results in a reduction of the water contact angle \cite{kova2009}.

As discussed in Section \ref{sec:exp_details}, both the porous and nonporous regions of each sample were exposed to the same processing and ambient environment. For this reason, the angles measured on the nonporous region represent the Young angles for the corresponding porous regions.

\begin{figure}
\includegraphics[scale=0.2]{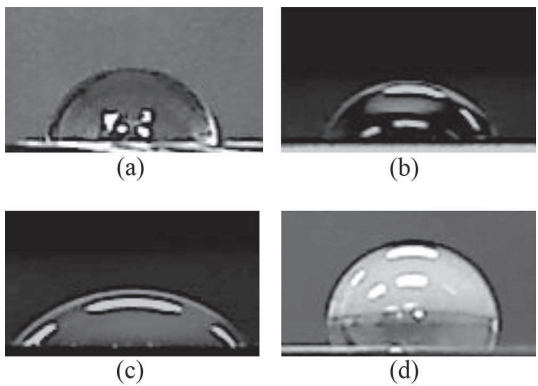}
\caption{Images of water drops on (a) microporous, (b) mesoporous, (c) as-prepared macroporous silicon, and (d) HF-dipped macroporous silicon.}
\label{fig:waterdrops}
\end{figure}

\subsection{Contact Angle: Porous Regions} \label{sec:cont_ang_porous}
Fig. 3 shows water drops on microporous, mesoporous and both as-prepared (i.e., post-KOH) and HF-dipped macroporous silicon films. As can be seen, the microporous and mesoporous films are hydrophilic with contact angles $< 90^\circ$. The same is
true for as-prepared macroporous silicon films. In contrast, the HF-dipped macroporous silicon films exhibit hydrophobicity. 

Table \ref{tab:cont_ang_films} compares the contact angles obtained in the present work to those obtained in previous studies \cite{kova2009,naga2009,liu2009,coff2013,ress2008,nova2011,spen2013}. The smaller contact angles measured on the microporous films are comparable to those measured in two earlier works \cite{kova2009,ress2008}. The relatively large range of angles (76$^\circ$-87$^\circ$) for the microporous films is likely due to partial oxidation \cite{will1974} due to exposure to ambient air both prior to vacuum storage and after removal for contact angle measurement. This type of porous silicon, in
fact, is known to oxidize very rapidly in air; for example, Ref. \cite{ress2008} reported a 70\% reduction in contact angle for microporous silicon after only one hour of air exposure.

Contact angles for the mesoporous films are consistent with those obtained on similar films prepared in our laboratory \cite{spen2013}. In contrast, other studies \cite{coff2013,nova2011} report contact angles $> 90^\circ$ for mesoporous silicon, indicating hydrophobicity. Again, the lower angles measured in the present study relative to these others
is likely due to some degree of oxidation due to air exposure.

The results for the as-prepared macroporous silicon films are similar to those for hydroxylated macroporous silicon \cite{liu2009}, suggesting that the hydrogen-terminated surface of our freshly-etched films have become oxidized to some extent as a result of the KOH etch. As discussed earlier, this etch is known to result in film oxidation. In contrast, HF-dipped macroporous films are hydrophobic, as reported in previous studies \cite{naga2009,ress2008}.

\begin{table}
\caption{Contact angles measured on micro-, meso-, and macroporous silicon films.} \label{tab:contact_angle_porous}
\begin{ruledtabular}
\begin{tabular}{ccc}
Type of Film & Reference &  Contact Angle \\ 
            & & (Deg) \\ \hline
\multirow{4}{*}{Microporous} & Pres. Work & 76-87 \\
                        & \cite{kova2009} & 75-105 \\
                        & \cite{ress2008} & 67 \\ \hline
\multirow{4}{*}{Mesoporous}  & Pres. Work & 67-79 \\
                            & \cite{spen2013} & 78 \\
                            & \cite{coff2013} & 104 \\
                            & \cite{nova2011} & 130 \\  \hline       
\multirow{4}{*}{Macroporous} & Pres. Work (As Prep) & 49-75 \\
                            & Pres. Work (HF Dip) & 99-111  \\
							& \cite{ress2008} & 110 \\
                            & \cite{naga2009} & 90-135 \\
                            & \cite{liu2009} & 60 \\ 
\end{tabular}
\end{ruledtabular}
\label{tab:cont_ang_films}
\end{table}

\subsection{Application of Wetting Models} \label{sec:app_wet_models}
Table \ref{tab:ave_cont_ang} shows that, for most films, the average water contact angles on the porous regions are greater than those on the corresponding nonporous regions, regardless of film type. Application of the Wenzel model (Eq. 1) to these films leads to $r_W \leq 1$, which is unphysical and demonstrates that the wetting is heterogeneous. For this reason, we apply the Cassie-Baxter model (Eq. 2) and estimate the liquid-solid surface contact fraction, $f$. The resulting values are shown in Table \ref{tab:ave_cont_ang}. The values obtained for the microporous and mesoporous films are large ($f \sim 0.9$), as are those for the as-prepared macroporous films ($f \sim 0.8$). These values are higher than expected given the porosity of the films. On the contrary, the $f$ values for the HF-dipped macroporous films are lower ($\sim 0.55$) and are in accord with those estimated from the film porosity. More specifically, because the areal and volumetric porosities of the macroporous films are approximately equal (see Section \ref{det_macro_film_porosity}), $f \approx 1 - P \approx 0.6$, where $P$ is $\sim 40$\% as determined from SEM image analysis.

It is interesting to note that there are two films (one microporous and one mesoporous, identified by asterisks in Table \ref{tab:ave_cont_ang}) for which the associated contact angle data gives physical values of Wenzel roughness parameter. Specifically, application of Eq. 1 yields $r_W$ values of approximately 4 and 2, for
the microporous and mesoporous films, respectively.  These values are comparable to previously reported values for these types of porous silicon \cite{ress2008,nova2011} but cannot be reconciled with roughness values calculated from the specific surface area and film thickness. In particular, the specific surface area for microporous silicon formed in a manner similar to that used in the present work \cite{hali1994} is approximately 500 m$^2$/cm$^3$ which, when multiplied by the $\sim 8$ $\mu$m pore depth, yields a Wenzel roughness parameter $r_W \sim 4000$. A similar calculation for mesoporous silicon \cite{heri1987}, with a specific surface area of 200 m$^2$/cm$^3$, yields $r_W \sim 1000$.  Moreover, the value of $r_W \sim 2$ for mesoporous silicon is not consistent with that obtained using a simple model which assumes cylindrical pores oriented with their axes perpendicular to the surface (a first approximation to the pore morphology of this type of porous silicon).  In this model, the total area accessible to the liquid is the internal pore surface area plus the area of remaining nonporous surface,
\begin{equation}
A_t = n(\pi r^2 + 2\pi r h) + (A - n\pi r^2),
\end{equation}
where $n$ is the total number of pores of radius $r$ and depth $h$, and $A$ is the total planar surface area.  Porosity is given by $P = n\pi r^2/A$, so that $r_W =
A_t/A$, or
\begin{equation}
r_W = \frac{2hP}{r} + 1.
\end{equation}
Here, $P \sim 60$\%, $h \sim 5$ $\mu$m and $r \sim 10$ nm, so that $r_W \sim 600$, the same order of magnitude as the value obtained from specific surface area considerations.

To gain further insight into the character of the wetting on porous silicon, a sample consisting of a microporous layer on a macroporous film (like that shown in Fig 1; see also Ref. \cite{peck2013}) was studied. The contact angle on the porous region of the as-prepared sample was found to be $83^\circ$, approximately 50\% greater than that on the adjoining nonporous region, a result inconsistent with the Wenzel model of wetting.  More interesting, however, is the similarity of this angle to the average contact angle measured on the single-layer microporous films ($\sim 82^\circ$) (see Table \ref{tab:ave_cont_ang}), suggesting that the liquid does not infiltrate the underlying macropores.

After dipping the microporous-on-macroporous sample in HF, the water contact angle on the porous region increased to $99^\circ$, about 20\% higher than that on the corresponding as-prepared structure. This large difference is likely due to removal of the top-lying microporous film by the HF. This is corroborated by the fact that a vigorous reaction was observed when the sample was immersed in HF and by the similarity of the contact angle to those measured on HF-dipped macroporous films (average $\sim 103^\circ$).  Moreover, the Cassie-Baxter $f$ value ($\sim 0.6$) is comparable to that for HF-dipped macroporous films ($f \sim 0.55$). This, coupled with the fact that the Wenzel model gives $r_W < 1$ for this film, again suggests
heterogeneous wetting.

\subsection{Possible Implications of Oxidation}
Collectively, the results above show that the predominant mode of wetting for the porous silicon films studied here is heterogeneous. Cassie-Baxter liquid-solid contact area fractions for many of the films are, however, larger than expected based on sample porosity. Moreover, the roughness ratios for the two films that exhibit possible Wenzel behaviour are orders of magnitude lower than those determined from specific surface area considerations.  These and other apparent contradictions of Section \ref{sec:app_wet_models} can be resolved by considering the effects of oxidation as discussed in Sections \ref{sec:cont_ang_nonporous} and \ref{sec:cont_ang_porous}.  For example, the large liquid-solid contact area fractions obtained for the microporous, mesoporous, and as-prepared macroporous films could result from oxdiation-induced pore narrowing, pore obstruction, and/or quasi-planarization of the porous topography. This explanation is corroborated by the results of several studies which show that oxidation can, in fact, result in significant pore narrowing \cite{pira2006,diaz2009,thom1998} and decreased surface roughness relative to unoxidized films \cite{diaz2009,char2007}. The same effects could also be responsible for the difference between the calculated and measured values of $r_W$ for the two films that exhibit possible Wenzel wetting. In such a case, the surface area accessible to the liquid could be dramatically reduced compared to that expected from specific surface area considerations, resulting in much smaller values of Wenzel roughness parameter. This explanation is supported by the contact angle results for the microporous, mesoporous, and even some of the as-prepared macroporous films as the relatively small difference ($\sim 10$\%) in angles measured on the porous and adjoining nonporous regions may be a consequence of oxidation.

\section{Conclusions}
Water contact angles have been measured on porous silicon films with pore diameters ranging from a few nanometers to about one micron. The results suggest that the predominant mode of wetting for these films is heterogeneous (either Cassie- Baxter-like or partial wetting of the pore interior).  While this study has focused on the wetting of porous silicon, the results obtained here may provide further insight into the nature of wetting in porous media in general.

\begin{acknowledgments}
GTA acknowledges the support of the Natural Sciences and Engineering Research Council of Canada (NSERC) (RGPIN-2015-04306).
\end{acknowledgments}

\bibliographystyle{apsrev4-2}
\bibliography{refs.bib}

\end{document}